\documentstyle[prl,aps]{revtex}

\begin{document}

\draft

\title{The Corkscrew Effect}

\author{John P. Ralston}

\address{Department of Physics and Astronomy, 
University of Kansas, Lawrence, KS 66044}

\author{Pankaj Jain}

\address{Physics Department, I. I. T., Kanpur, India 208016}

\author{Borge Nodland}

\address{Rochester Theory Center for Optical Science and
Engineering, University of Rochester, Rochester, NY 14627}

\date{Submitted to Physical Review Letters (1997)}

\maketitle

\begin{abstract}
We discuss a new mechanism which could cause a rotation of polarization
of electromagnetic waves due to magnetic fields on cosmological scales.
The effect is due to the geometrical phase of Pancharatnam and Berry,
and causes a corkscrew twisting of the plane of polarization. The new
effect represents an additional tool that allows possible intergalactic
and cosmological magnetic fields to be studied using radio
propagation.
\end{abstract}


Many studies have been done to find a cosmological component to
magnetic fields. A common method looks for systematic dependence of the
Faraday rotation measures ($\text{RM}$) with increasing redshift $z$.
For a radio wave of wavelength $\lambda$ and polarization angle
$\vartheta$, Faraday rotation\cite{one} predicts $\vartheta = \text{RM}
\lambda^2 + \chi$, where $\text{RM} = [e^3/(16 \pi^3 m_e
\varepsilon_0c^3)] \int^z B_z n_e \, dz$ for propagation through a
magnetic field with component $B_z$ along the propagation path; $n_e$
is the electron number density of the plasma, and $\chi$ is the angle
of emission.

Observation of the Faraday effect measures integrated magnetic plasma
parameters along the line of sight. Although extremely stringent bounds
have occasionally been stated \cite{two}, the methodology has its
limitations, inasmuch as reversal of the magnetic field will reverse
the rotation measure. This is not always discussed.  When possible
magnetic field reversals are taken into account, the observed rotation
measures are not capable of bounding local magnitudes of the fields:
magnetic fields larger than some bounds could well exist if the fields
go through some reversals of sign along the way.

Rotation of the plane of polarization is usually attributed to
birefringence.  Here, we present a new physical phenomenon, which we
call the ``corkscrew effect,''  which could also create a rotation of
polarization. The underlying observation is that any phase shift of the
positive relative to the negative helicity amplitudes of the wave will
rotate a linear polarization. The corkscrew effect occurs for
propagation with angular frequency $\omega$ in an  electromagnetic wave
equation of the form

\begin{equation}
[k^2\delta_{ij}  - k_i k_j - \omega^2 \varepsilon_{ij}(\omega)] E_j = 0;
\;\;\;
\varepsilon_{ij}(\omega) =
\varepsilon_{ij}^S(\omega) + i \varepsilon^A_{ij} (\omega),
\label{eq1}
\end{equation}
where $\varepsilon^S(\omega)$ is real and symmetric, and $\varepsilon^A
(\omega)$ is non--zero, real and antisymmetric; the full dielectric
constant is $\varepsilon(\omega)$ Hermitian (no
absorption)\cite{three}. We are interested in circumstances where the
propagating transverse modes are very nearly degenerate (free space
$\varepsilon_{ij} = \delta_{ij}$ in our units), with small
perturbations lifting the degeneracy. The system then becomes a
two-state evolution problem, analogous to two-state mixing problems
encountered in quantum mechanics, with $\varepsilon(\omega)$ with
$\epsilon$ representing parameters that can vary slowly along the path
of propagation. Where slow parameter variations are an issue, naive
inferences about wave propagation can be wrong. For example, if
$k_1(z)$ and $k_2(z)$ are the wave numbers from diagonalizing the
dispersion relation for fixed parameters and fixed $\omega$, one does
generally not find the phase of a wave from the expected eikonal
factors $\int^z i \{k_1(z), k_2(z)\} \, dz $. This is due to the
geometric phase phenomenon from parallel transport in the adiabatic
limit, as described for visible--light optics by Pancharatnam and Berry
\cite{four}.  We focus here on this purely geometrical effect, noting
for later study that strong dynamical effects can also occur -- such as
seen in resonant magnetic moment transitions, or resonant neutrino
oscillations -- but in this case resonances between coupled radio waves
traveling in a tenuous medium.  Interestingly, we find that sudden
resonant transitions do nothing observable, while partially resonant
transitions and the adiabatic limit produce very interesting results
indeed.

Application of electrodynamics to a magnetized neutral plasma (with an
external magnetic field $\vec{B} = B \widehat{B}$) gives
$\varepsilon_{ij} = \varepsilon_p \delta_{ij} - i \omega_B
\varepsilon_{ijk} \widehat{B}_k (\partial \varepsilon_p /\partial
\omega)$, with $\varepsilon_p = 1 - \omega_p^2/\omega^2$, where
$\omega_p$ is the plasma frequency, given approximately by $\omega_p^2
= 4 \pi e^2 n_e /m_e$, and $\omega_B$ is the cyclotron frequency
$\omega_B = e B / m_e$.  Small corrections of order $k T / m_e$ for a
non-relativistic hot magnetized plasma \cite{three} have no effect on
the discussion. A common method of solution \cite{one} assumes that
the magnetic field and the wave vector are parallel. One readily
diagonalizes the dielectric tensor for this case, which has transverse
eigenstates of circular polarization $\vec{E}_\pm$ and wavenumbers
$k_+$ and $k_-$ for the same $\omega$.  It is tempting to focus on
the part of $\vec{B}$ parallel to $\vec{k}$, while treating other
components ($\vec{B}_\perp$) as a perturbation.  Then, goes the usual
(but faulty) argument, the total rotation of polarization when the
parameters vary along the line of propagation would be given by
integrating $(1/2)(k_+-k_-)$ along the path, resulting in the
astronomers' rule $\Delta \theta = \text{RM} \lambda^2 $ for $\omega \gg
\omega_p, \omega_B$ \cite{one}. This seems rigorous enough in
astrophysical circumstances, where $\omega \approx 10 \text{MHz} - 1
\text{GHz}$, $n_e \approx 10^{-6} \text{cm}^{-3} - 10^{-3}
\text{cm}^{-3}$, and $B \approx 10^{-9} \text{G} - 10^{-5} \text{G}$,
and the approximation $\omega \gg \omega_B, \omega_p$ is satisfied by
several orders of magnitude.

Our point of departure comes from noting that the propagation problem
for $\vec{B} = 0$ is utterly degenerate, creating conditions for a
corkscrew effect: the direction of $\vec{B}$ lifts the degeneracy and
selects unique eigenstates. Treating the transverse parts of $\vec{B}$
as a perturbation will not do, because the effects of the perturbation
are the same order as the splitting of eigenvalues. We solved the
problem exactly for the two transverse eigenstates of $D_i  =
\varepsilon_{ij} E_j$ used in the adiabatic limit. One needs the
eigenvalues of the inverse dielectric tensor in the $2 \times 2$
subspace orthogonal to $\vec{k}$, which are given by
$\varepsilon^{-1}_\pm = \frac{\varepsilon_p}{\varepsilon_p^2-b^2} -
\frac{b^2 \sin^2\vartheta_{kB} }{2\varepsilon_p(\varepsilon_p^2 - b^2)}
\pm \frac{b \sqrt{\cos^2 \vartheta_{kB} + b^2 \sin^4 \vartheta_{kB} }
}{2 \varepsilon_p (\varepsilon_p^2 - b^2)}$. The magnetic field
direction has been represented by polar coordinates ($\vartheta_{kB},
\phi_{kB}$) in the coordinate system where $\vec{k}$ is along the $z$
axis; $b = \omega_p^2 \omega_B/\omega^3 \ll 1$ is a dimensionless small
parameter measuring the strength of the $B$ field. For finite $b$, the
system is non--degenerate for any direction of $\vec{B}$. As
$\cos(\vartheta_{kB}) \rightarrow 0$, the splitting of the eigenvalues
reaches a non-zero minimum, exhibiting an avoided level crossing.

The exact solution also reveals that the polarization eigenvectors
deviate strongly from circular polarization when $\vec{k}$ and
$\vec{B}$ are nearly perpendicular, or, more precisely, when
$\cos\vartheta_{kB}$ is of order $b$. A standard way to express the
polarization of a radiation eigenvector uses the Poincar\'{e} sphere
\cite{one}. In the helicity basis, any normalized complex two--vector
$|s\rangle$ can be associated with a $2 \times 2$ traceless,
determinant $-1$, Hermitian matrix $P$, which satisfies $P |s\rangle =
|s\rangle$.  If we write

\begin{equation}
P = \left(
\begin{array}{cc}
\cos \vartheta_p & \sin \vartheta_p e^{-i \phi_p} \\
\sin \vartheta_p e^{i \phi_p} & -\cos\vartheta_p
\end{array}
\right),
\label{eq2}
\end{equation}
then the angles $(\vartheta_p, \phi_p)$ define the polarization state
of the vector. The positive and negative circular polarizations are
represented by the top and bottom poles of the sphere; linear
polarizations are represented by points on the equator.  The solution
for the eigenstate of definite frequency $d_1$ is found to be

\begin{equation} 
\tan(\vartheta_p) = 
b \sin^2(\vartheta_{kB}) /\cos(\vartheta_{kB}); 
\; \; \phi_p = -2 \phi_{kB}, \label{eq3}
\end{equation} 
with $d_2$ following from orthogonality. To understand the physical
effects, from (\ref{eq3}), we see that the eigenvectors ``follow'' the
magnetic field, but in a highly non--linear fashion. For $b \ll 1$, and
$|\cos(\vartheta_{kB})|$ not too small, the eigenvectors are
nearly circularly polarized, living on tiny circles with $\phi_p = -2
\phi_{kB}$ near the top or bottom of the Poincar\'{e} sphere. (The
double--valuedness of this map is a reflection of the projective
character of polarization).

Something very dramatic happens when $\vec{B}$ rotates transverse to
$\vec{k}$.  The dominance of circular polarization is upset, even for
non--zero $b$ arbitrarily small, as the components of $\vec{B}$
transverse to $\vec{k}$ receive all the responsibility for lifting the
degeneracy. The position on the Poincar\'{e} sphere jumps from its
azimuthal position near the pole to the equator just as $B_z$ crosses
zero. This will occur in any circumstance (see Fig. 1) when the sign of
the component of the magnetic field in the $z$--direction changes
(a ``$B_z$ reversal'').

It is convenient to consider the case where the magnetic field
direction traverses a closed orbit along the path. We present the cases
shown in Fig. 1(a), where the $\vec{B}$ rotates in a spiral about
$\vec{k}$ without reversing, compared to another case [Fig. 1(b)] where
$B_z$ reverses while undergoing an azimuthal twist. The corresponding
trajectories of the polarization on the Poincar\'{e} sphere are also
shown. Although the system parameters return to their initial
configuration after the cycle, the eigenstates pick up a geometrical
phase, with magnitude on the positive ($E_+$) eigenvector given by the
solid angle $\Omega$ traversed on the Poincar\'e sphere. The phase on
the negative ($E_-$) eigenvector is easily shown to be the same with
opposite sign, so that the phase difference of the two circular
polarizations is twice the solid angle swept out. This phase
difference, which is just the angle by which the plane of polarization
is rotated, is entirely of geometrical origin, and is in addition to
the eikonal phase $\exp[i \int^z k_j(z) \, dz]$.

The geometrical character of the process can be intuitively visualized
by considering the effects on a two--dimensional Cartesian coordinate
frame (representing the set of polarizations) erected tangent to the
sphere. As the coordinate frame is dragged along parallel to itself,
its orientation accumulates information on the curvature of the state
space. For the no $B_z$ reversal case [the spiral field of Fig. 1(a)],
the solid angle $\Omega$ is of order $b^2$. A phase shift of order
$b^2$ would be quite small, and give a rotation of polarization going
like $\lambda^6$. On the other hand, suppose the field $B_z$ reverses,
dragging the tangent frame from top to bottom of the sphere.  At the
bottom, the frame (oriented at the angle $\phi_p$) undergoes a twist by
the azimuthal angle $-2\Delta \phi_{kB}$, then $B_z$ reverses again,
followed by $+2\Delta\phi_{kB}$. The net phase shift is $\Omega=4\Delta
\phi_{kB}$. The $\phi_B$ dependence of $\phi_p$ is crucial. Of course it
is not necessary for $b$ to be constant with $z$. Nor is a closed
circuit necessary. When the slow parameters $p^k$ vary arbitarily, the
corkscrew effect gives a rotation of polarization by an angle $\phi =
2\int^{p(z)} dp^k \cdot A_p^k(z)]$, which is of order unity and which is
determined solely by changes of directions of $\vec{B}$. Here $A^k_p =
(1- cos\vartheta_p)\partial^k\phi_p$, where $\partial^k$ is the
derivative with respect to the parameters $[b(z),
\cos\vartheta_{kB}(z),\phi_{kB}(z)]$, or, more directly,
$(\vartheta_p,\phi_p)$. (In Berry's convention \cite{four}, the phases
are already attached to eigenstates.)  If there is more than one $B_z$
reversal along the trajectory, the various contributions track the
history of the polarization states along the trajectory.

Borrowing quantum terminology, we have just examined the adiabatic
approach to resonance where a phase shift of geometrical origin
occurs.  Under adiabatic conditions, the ``probability'' (amplitude
squared) of each eigenstate is conserved, while the phase--physics is
non--trivial.  The adiabatic approximation applies when $p^{-1}(z) d
p(z)/dz \ll k b$.  This means that parameters vary over a length $L$
such that $\omega^2_p\omega_B L / \omega^2 > 1$, that is, that the
wavelength is not too short. Comparing the scales, this is also the
circumstances where Faraday rotation is non--negligible. Numerous
interesting phenomena also occur outside of the adiabatic limit. In
non--adiabatic resonance transitions the probability can hop from one
state to another, again generating non- trivial dynamical phases. In
the extreme case of the ``sudden'' approximation the parameters vary so
rapidly (on the distance scale of the wavelength splitting) that the
system jumps abruptly from old eigenstates to new eigenstates based
only on the overlap (dot--product) of the states. For the magnetized
plasma, this is the limit of $\omega \rightarrow \infty$. With $b$ (the
dimensionless level splitting parameter) going like $1/\omega^3$, the
splitting becomes so small that any physical variation of parameters
becomes ``sudden.'' Transitions occur on the basis of maximum overlap,
and when the eigenstates switch, the probability switches
correspondingly. This answers the paradox that the $\vec{B}$--field
actually does lift the degeneracy of a gamma ray, but with negligible
results: the corkscrew effect, to be sizable, occurs when the ordinary
Faraday rotation is sizable.  One cannot avoid the conclusion that
under a wide variety of circumstances, a $B_z$--reversal will produce
non--trivial phase shifts between the helicity states and a substantial
frequency--independent rotation of polarization.

Let us now turn to astrophysical implications of these results.
Current knowledge of intergalactic magnetic fields is scanty.  Even the
basic origin of established galactic fields is unknown. There are two
major theories invoking internal dynamos or pre--existing primordial
magnetic fields \cite{five}. Indications of a coherent field coming
from excess $\text{RM}$'s of extragalactic sources has been claimed,
but the interpretation of the $\text{RM}$ data is ambiguous \cite{six}.
The existence of magnetic fields in galaxies at large redshifts has
been argued to be a challenge to galaxy dynamo theory \cite{seven}.

Recently, an indication of anisotropy in the propagation of radio waves
on cosmological distances was reported \cite{eight}. After the Faraday
effect was subtracted, a signed difference of angles called $\beta$ was
observed to behave as $\beta=r/2\Lambda_s \cos(\gamma)$ [In the
analysis of \cite{eight}, care was used to employ a variable ($\beta$)
which properly reflects the projective nature of polarization
observables.  Combinations such as $\chi-\psi$, $| \chi-\psi |$, and
the acute angle between $\chi$ and $\psi$ cannot serve to detect
cosmological birefringence.] Surprisingly, the direction of the
anisotropy axis extracted in \cite{eight} coincides within errors with
the direction of the dipole component of the cosmic microwave
background \cite{nine}, perhaps indicating that some type of
intergalactic or cosmological medium may be at work. The possibility of
systematic bias in the data was discussed, with emphasis on corrections
made by the observers for our own galaxy.  Studying the internal
structure of the data, it was noticed \cite{eight} that strong
correlations were observed for about half the data for galaxies with $z
\geq 0.3$, while no significant correlation was observed for the half
of the data with $z<0.2-0.3$. If something not taken into account were
due to our own galaxy, it should affect all data for the extragalactic
sources in the same way.  As reported in \cite{eight}, unexpected
physics including the corkscrew effect that might occur in our local
neighborhood cannot account for the observations.

Whether a magnetic field of cosmological extent could explain the same
correlation is a very interesting question. The turning--on of the
correlation for $z > 0.3$ may be an indication that a system enormously
larger than our local group trails a magnetic field or some other
condensate of huge size. Galaxy density distribution studies tend to
support the presence of coherent structure over vast scales, although
this is controversial \cite{ten}. Complementary information might come
from measurements of the cosmic microwave background polarization
affected by primordial magnetic fields existing long before galaxy
formation \cite{eleven}.  Moreover, there could be a wide range of
interpretations for the results of \cite{eight}, including anisotropic
cosmologies of general relativity, an axion--like condensate domain
wall, universal rotation, spontaneous rotational symmetry breaking, or
antisymmetric degrees of freedom in gravitational theory \cite{twelve}.
We cannot begin to answer all the questions here. A systematic approach
using all the tools available, and orders of magnitude more data, would
be able to make a significant scientific contribution.

While focusing here on the parameters affecting linear polarization,
there is clearly a scientific gold mine in the accumulation of all
radio polarization observables which have never been interpreted taking
into account the effects we discuss. It is very important to have data
at different frequencies, in order to exploit the full potential of
observations. The corkscrew effect might serve as a valuable discovery
tool when larger data sets become available, with many sources examined
over a range of radio frequencies, supplemented by optical and
ultraviolet polarization data.

Doug McKay made a helpful suggestion. PJ thanks the High Energy Group
of ICTP (Trieste) for hospitality during completion of this work under
the Associate Membership Programme. DOE Grant DE--FG02--85ER40214, the
K${}^*$STAR program under the Kansas Institute for Theoretical and
Computational Science, Grant DAE/PHY/96152, and NSF Grant PHY94--15583
provided support.

\begin{figure}
\caption{
Cartoons illustrating a phase change from slow variations in the
direction of $\vec{B}$ along the path of the wave. (a) $\vec{B}$ varies
in a spiral, producing a tiny circular orbit on the Poincar\'{e} sphere;
(b) $B_z$ reverses while $B_\phi$ twists, producing a flip and twist on
the sphere. Corresponding points (1--5), with the configurations at 5
equal to 1, are shown.}
\label{fig1}
\end{figure}

\end{document}